\documentclass[12pt,dvips]{article}

\usepackage{rotating,amsmath}
\usepackage{epsfig}
\usepackage{color}
\usepackage{cite}

\usepackage{hhline}
\usepackage{multirow}

\newcommand{\lsim}{\raisebox{-0.13cm}{~\shortstack{$<$ \\[-0.07cm] $\sim$}}~}
\newcommand{\gsim}{\raisebox{-0.13cm}{~\shortstack{$>$ \\[-0.07cm] $\sim$}}~}

\begin{document}

\def\thefootnote{\fnsymbol{footnote}}

\begin{flushright}
{\tt  KIAS Preprint P11017}\\
April 2011
\end{flushright}

\vspace{1cm}
\begin{center}
{\bf {\Large Possible Common Origin of \\ 
the Top Forward-backward Asymmetry \\[0.2cm] 
and the CDF Dijet Resonance}}
\end{center}

\medskip

\begin{center}{\large
Dong-Won~Jung$^{a,b}$,
P.~Ko$^{c}$ and
Jae~Sik~Lee$^{a,b}$}
\end{center}

\begin{center}
{\em $^a$Department of Physics, National Tsing Hua University,
Hsinchu, Taiwan 300}\\[0.2cm] 
{\em $^b$Physics Division, National Center for Theoretical Sciences, 
Hsinchu, Taiwan 300}\\[0.2cm]
{\em $^c$School of Physics, KIAS, Seoul 130-722, Korea}
\end{center}

\bigskip\bigskip

\centerline{\bf ABSTRACT}
\noindent  
A color-singlet neutral vector boson is considered as a possible
common origin of the top forward-backward asymmetry and the
CDF dijet resonance. We identify chiral and flavor structures of
the couplings of this new vector boson to the standard model quarks
for which one could accommodate both data.
We also demonstrate that non-vanishing observables 
involving longitudinal top polarizations can provide
useful criteria for the possible existence of parity violating
new physics in the $q\bar{q} \to t\bar{t}$ process, and
discrimination between the flavor-conserving
and flavor-violating cases.



\newpage

\section{Introduction}
\label{sec:intro}

Top forward-backward (FB) asymmetry has been exciting subject, since the data showed
deviation from the SM predictions at the level of 2-3 $\sigma$~\cite{Aaltonen:2011kc}.
There have been a lot of study on this subject. It is still premature what kind of
new physics could be responsible for the observed deviations. If the 
FB asymmetry
can be measured as functions of $M_{t\bar{t}}$ and $\Delta  y \equiv y_t - y_{\bar{t}}$
with nontrivial structures in them, it could provide more informations on the
underlying physics. Also the measurements of the FB spin-spin correlations in the
$t\bar{t}$ production and (anti)top longitudinal polarization could provide more
informations on the underlying physics behind the observed top FB asymmetry
\cite{Jung:2009pi,Jung:2010yn}.

Recently the CDF Collaboration reported another interesting data 
on $Wjj$ channel~\cite{Aaltonen:2011mk}.
The result is that a broad peak 
in the $120-160$ GeV dijet invariant mass range,
with the estimated production cross section $\sim 4$ pb.
There is no evidence that these dijets are $b$-flavored, and no clear resonance
structure in $Wjj$ invariant mass spectrum.
A number of papers stimulated by this peak have appeared discussing
it in various contexts~\cite{Buckley:2011vc,Yu:2011cw,Eichten:2011sh,Kilic:2011sr,
Cheung:2011zt,AguilarSaavedra:2011zy,He:2011ss,Wang:2011ta,Sato:2011ui,
Nelson:2011us,Anchordoqui:2011ag,Dobrescu:2011px,Fodor:2011tu,Jung:2011ua,
Buckley:2011vs,Zhu:2011ww,Sullivan:2011hu,Ko:2011ns,Plehn:2011nx}.
%
The D$0$ Collaboration also reported the 
study of the dijet invariant mass distribution
in $p\bar p \to W (\to l\nu) + jj$ final states~\cite{Abazov:2011af}.
They found no evidence for anomalous resonant dijet production and set
a 95 \% C.L. limit of 1.9 pb on the cross section for 
the dijet invariant mass $m_{jj}=145$ GeV.
In this work,
without clear understanding of the discrepancy between the two experiments,
we postulate a new particle with the production cross section 
of $1 - 4$ pb at the Tevatron.
We further assume that it couples dominantly to 
the quarks in order to evade strong constraints from Drell Yan process.
%
%
%

It would be interesting to ask if one can explain both top FB asymmetry and
the dijet resonance in the $Wjj$ channel by introducing a single new particle.
It is our purpose to answer this question using a neutral color-singlet vector 
boson $V_\mu$, assuming the CDF dijet peak  is due to 
$p\bar{p} \rightarrow W V \rightarrow (l\nu)(jj)$.
Let us  consider the following New Physics (NP) interactions of $V_\mu$: 
\begin{eqnarray}
{\cal L}_{\rm NP}=
-\,g_s\,\sum_{q=u,d,t} \overline{q}\,\gamma^\mu \left(g^q_L P_L +g^q_R P_R \right) q\, V_\mu
-\,\left[g_s \overline{u}\gamma^\mu \left(\tilde{g}^t_L P_L +
\tilde{g}^t_R P_R \right) t \, V_\mu + h.c. \right]\,.
\label{eq:NP}
\end{eqnarray}
The first and second terms describe the flavor-conserving (FC) and flavor-violating (FV)
interactions, respectively. We are including the interactions of $V_\mu$ with 
the first-generation quarks and top quarks, since our interest is in 
$p\bar{p} \rightarrow t \bar{t}, ~ W V$.
We are using the strong coupling constant $g_s$ for the overall normalization of
the FC $g^q_{L,R}$ and FV $\tilde{g}^t_{L,R}$ couplings for consistency with the 
model-independent studies of the Tevatron forward-backward asymmetry of 
top quark ($A_{\rm FB}$)~\cite{Jung:2009pi,Jung:2010yn}.

Within this framework, we study hadroproductions of $WV$ and $V$ as well as
$t\bar{t}$ production and the top FB asymmetry at the Tevatron, 
and identify the chiral and the flavor structures of couplings that 
can accommodate both top FB asymmetry and the CDF dijet excess.

\section{Hadroproductions of $W V$ and $V$}
\label{sec:production}

The $V$ production associated with $W^\pm$ 
at hadron colliders occurs via 
the exchanges of the $u$, $d$, and $t$ quarks. 
We denote the $WV$ production cross section as the sum
\begin{equation}
\sigma_{LO}({\rm had}_1{\rm had}_2\to W^-V)=
(g^u_L)^2\, \sigma^u_L \ + \ 
(g^d_L)^2\, \sigma^d_L \ + \ 
(g^u_L g^d_L)\, \sigma^{ud}_L \ + \ 
|\tilde{g}^t_L|^2\, \sigma^t_L \ + \ 
|\tilde{g}^t_R|^2\, \sigma^t_R \,,
\end{equation}
where 
\begin{eqnarray}
\sigma^X_{L,R} & = &
\int_{\sqrt{\hat{s}}_{\rm min}}^{\sqrt{\hat{s}}_{\rm max}}\,{\rm d}\sqrt{\hat{s}}
\int_{\hat{t}_{\rm min}}^{\hat{t}_{\rm max}}\,{\rm d}\hat{t}
\int_\tau^1\,{\rm d}x
\\[0.2cm]
&\times & \left[
\frac{\tau}{x}f_{\rm had_1}^{D}(x,Q_F) f_{\rm
had_2}^{\bar{u}}\left(\frac{\tau}{x},Q_F\right)+\frac{\tau}{x}f_{\rm had_1}^{\bar{u}}(x,Q_F)
f_{\rm had_2}^{D}\left(\frac{\tau}{x},Q_F\right)
\right]\,
\left(\frac{2}{\sqrt{\hat{s}}}
\frac{{\rm d}\hat\sigma^X_{L,R}}{{\rm d}\hat{t}}\right) \nonumber
\end{eqnarray}
where $(X,D)=(u,d),(d,d),(ud,d), (t,b)$ and $\tau=\hat{s}/s$
with $s$ being the centre-of-mass the colliding hadrons. And
$\hat{s}=(p_1+p_2)^2=(k_1+k_2)^2$,
$\hat{t}=(p_1-k_1)^2=(k_2-p_2)^2$, and
$\hat{u}=(p_1-k_2)^2=(k_1-p_2)^2$ with $p_1$ and $p_2$ for the momenta of the initial
$D$ and $\bar{u}$ quarks, respectively, and
$k_1$ and $k_2$ for those of the outgoing $W^-$ and $V$ vector bosons, respectively.
The symbol
$f_{\rm had_i}^u(x,Q_F)$, for example, is for the $u$-quark distribution
function in the hadron ${\rm had_i}$. The kinematic range of $\hat{t}$ is given by
\begin{eqnarray}
{\hat{t}_{\rm max\,,min}}=\frac{1}{2}\,\left(
m_V^2+m_W^2-\hat{s}\right) \ \pm \
\frac{\hat{s}}{2}\,\lambda^{1/2}
\end{eqnarray}
with $\lambda=1+(m_V^2/\hat{s}-m_W^2/\hat{s})^2-2m_V^2/\hat{s}-2m_W^2/\hat{s}$.
The range of the variable $\hat{s}$ can be determined by requiring
$0<\lambda<1$ and $x>\tau$. Also note the relation
$\hat{s}+\hat{t}+\hat{u}=m_V^2+m_W^2$.
The partonic-level cross sections are given by
\begin{eqnarray}
\frac{{\rm d}\hat{\sigma}^u_L}{{\rm d}\hat{t}}&=& \frac{\pi\alpha\alpha_S}{6s_W^2 \hat{s}^2}
\frac{\left(\hat{u}\hat{t}-m_V^2m_W^2\right)+\hat{t}^2E\left(\hat{s},\hat{t},\hat{u}\right)}{\hat{t}^2},
\nonumber \\
\frac{{\rm d}\hat{\sigma}^d_L}{{\rm d}\hat{t}}&=& \frac{\pi\alpha\alpha_S}{6s_W^2 \hat{s}^2}
\frac{\left(\hat{u}\hat{t}-m_V^2m_W^2\right)+\hat{u}^2E\left(\hat{s},\hat{t},\hat{u}\right)}{\hat{u}^2},
\nonumber \\
\frac{{\rm d}\hat{\sigma}^{ud}_L}{{\rm d}\hat{t}}&=& \frac{\pi\alpha\alpha_S}{6s_W^2 \hat{s}^2}
\left[ 2 \hat{s}\frac{m_V^2+m_W^2}{\hat{u}\hat{t}}-2E\left(\hat{s},\hat{t},\hat{u}\right)
\right], \nonumber\\
\frac{{\rm d}\hat{\sigma}^t_L}{{\rm d}\hat{t}}&=& \frac{\pi\alpha\alpha_S}{6s_W^2 \hat{s}^2}
\frac{\left(\hat{u}\hat{t}-m_V^2m_W^2\right)+\hat{t}^2E\left(\hat{s},\hat{t},\hat{u}\right)}
     {\left|\hat{t}-m_t^2+im_t\Gamma_t\right|^2},
 \nonumber \\
\frac{{\rm d}\hat{\sigma}^t_R}{{\rm d}\hat{t}}&=& \frac{\pi\alpha\alpha_S}{6s_W^2 \hat{s}^2}
\frac{m_t^2\left[\hat{s}+\hat{t}F\left(\hat{s},\hat{t},\hat{u}\right)\right]}{\left|\hat{t}-m_t^2+im_t\Gamma_t\right|^2},
\end{eqnarray}
where
\begin{eqnarray}
E(\hat{s},\hat{t},\hat{u}) & =& 
\frac{1}{4}\left(\frac{\hat{u}\hat{t}}{m_V^2m_W^2}-1\right)+
\frac{1}{2}\frac{(m_V^2+m_W^2)\,\hat{s}}{m_V^2m_W^2}
\nonumber \\
F(\hat{s},\hat{t},\hat{u}) & =& 
\frac{\hat{s}\hat{t}}{4m_V^2m_W^2}+
\frac{1}{2}\frac{(m_V^2+m_W^2)(\hat{u}\hat{t}-m_V^2m_W^2)}{m_V^2m_W^2\,\hat{t}}
\end{eqnarray}
We verified 
our expressions agree with those given in Ref.~\cite{Brown:1979ux} 
for $\hat\sigma^{u,d,ud}$.  
Note the non-vanishing new contribution 
$\hat\sigma^t_{L,R}$ 
due to the heavy top-quark exchange.
One may obtain similar expressions for 
$\sigma_{LO}({\rm had}_1{\rm had}_2\to W^+V)$.

\begin{figure}[!t]
\begin{center}
{\epsfig{figure=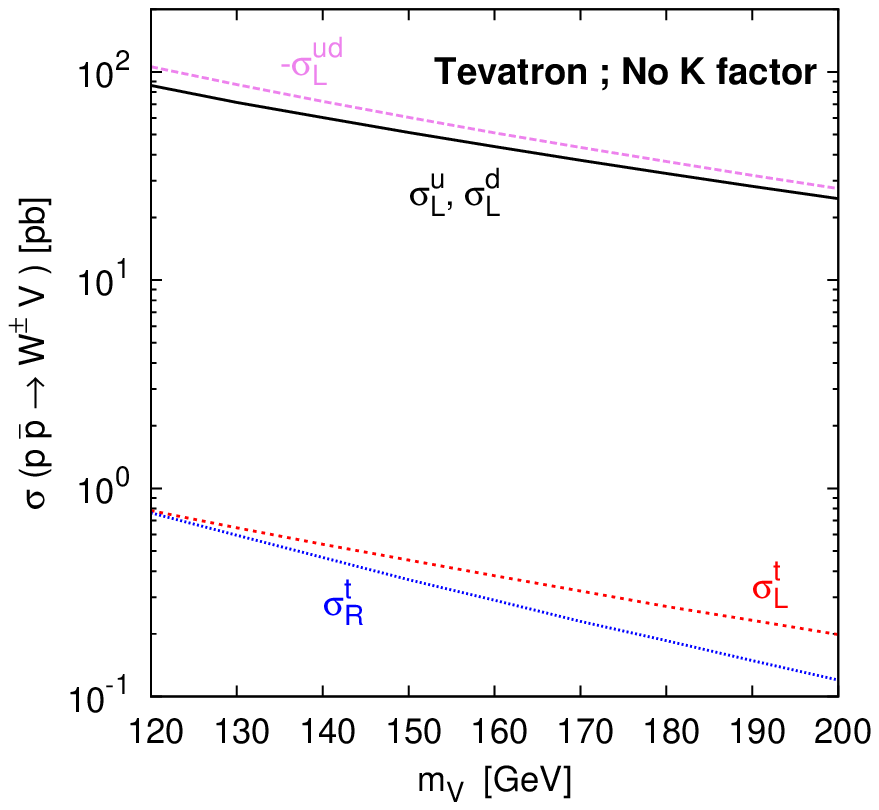,height=7.0cm,width=9.5cm}}
\hspace{-3.0cm}
{\epsfig{figure=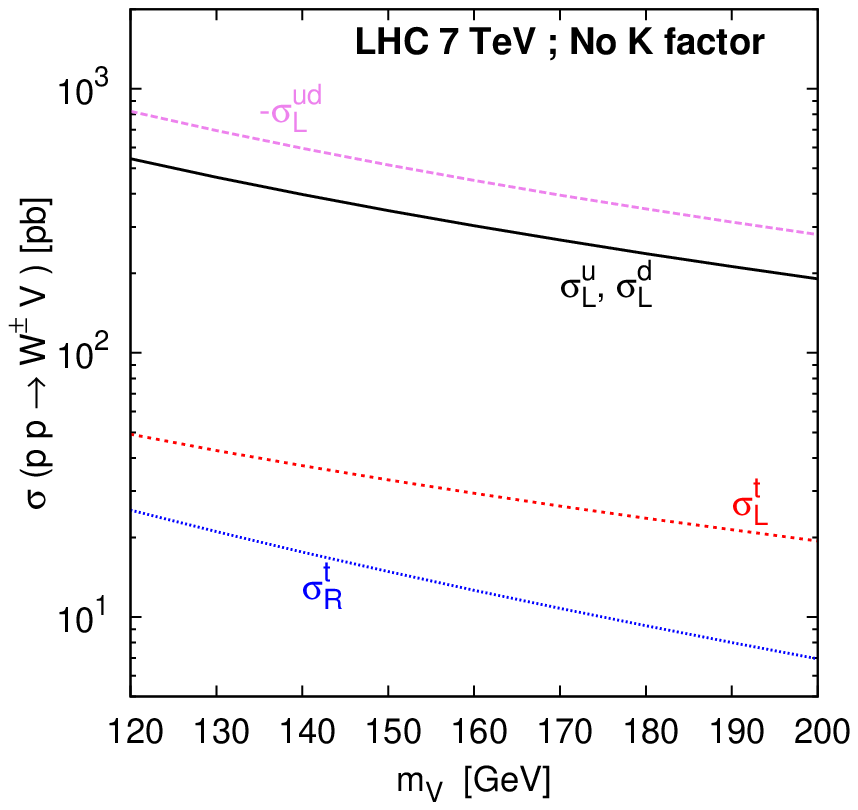,height=7.0cm,width=9.5cm}}
\end{center}
\vspace{-1.0cm}
\caption{\it The cross sections 
$\sigma^{u}_L$, $\sigma^d_L$, $-\sigma^{ud}_L$, and
$\sigma^t_{L,R}$ as functions of $m_V$ for
$\sigma_{LO}(p\bar{p}\to W^\pm V)$ at the Tevatron (left)
and at the LHC with 7 TeV (right).
}
\label{fig:qqwv}
\end{figure}

Fig.~\ref{fig:qqwv} shows the cross sections at the Tevatron
and at the LHC with 7 TeV.
The cross section $\sigma_{LO}(p\bar{p}\to W^\pm V) \sim 1$ pb can be 
easily achieved when $({g}^{u,d}_{L})^2 \sim 1/50$ taking $m_V = 150$ GeV.
It also can be achieved even for the top-quark exchange case 
but with a bit large couplings $|\tilde{g}^{t}_{L,R}|^2 \sim 3$ which might
be constrained by the $t\bar{t}$ production cross section.
%

In this work, we have taken account of the constraints from
the non-observation of the $s$-channel $V$ production
by UA2 collaboration~\cite{Alitti:1990kw}.
The resonant cross section may be given by
\begin{equation}
\sigma_{LO}(p\bar{p}\to V) = \sum_{q=u,d}\,
\sigma_{LO}(q\bar{q} \to V) = \frac{2\pi^2\alpha_S}{3m_V^2}\,\sum_{q=u,d}\,
\left[(g^q_L)^2+(g^q_R)^2\right]\,
\left(\tau\frac{{\rm d}{\cal L}^{q\bar{q}}}{{\rm d}\tau}\right)\,.
\end{equation}
Here $\tau=m_V^2/s$ with $\sqrt{s} =630$ GeV and
\begin{equation}
\tau\frac{{\rm d}{\cal L}^{q\bar{q}}}{{\rm d}\tau}=
\int_\tau^1\,{\rm d}x\,\left[
\frac{\tau}{x}f_p^q(x,Q_F) f_{\bar{p}}^{\bar{q}}\left(\frac{\tau}{x},Q_F\right)+
\frac{\tau}{x}f_p^{\bar{q}}(x,Q_F) f_{\bar{p}}^q\left(\frac{\tau}{x},Q_F\right)\right]\,.
\end{equation}
%

\section{Top FB asymmetry and polarization observables}
\label{sec:afb}
The forward-backward asymmetry $A_{\rm FB}$ of the top quark is
one of the interesting observables related with top quark.
The most recent measurement in the $t\bar{t}$ rest frame is
\cite{cdf2010}
\begin{eqnarray}
A_{\rm FB} & \equiv & \frac{N_t ( \cos\theta \geq 0) - N_{\bar{t}}
( \cos\theta \geq 0 )}{N_t ( \cos\theta \geq 0) + N_{\bar{t}}
( \cos\theta \geq 0 )} = 
(0.158 \pm 0.072  \pm 0.017)
\end{eqnarray}
with $\theta$ being the polar angle of the top quark with
respect to the incoming proton in the $t\bar{t}$ rest frame.
Within the SM, this asymmetry vanishes at leading order
in  QCD because of $C$ symmetry.
At next-to-leading order [$O(\alpha_s^3)$],
a nonzero $A_{\rm FB}$  can develop 
with the prediction $A_{\rm FB}\sim 0.078$ \cite{Antunano:2007da}.

Another interesting observable which is sensitive to the chiral
structure of new physics affecting $q\bar{q} \rightarrow t\bar{t}$
is the top quark spin-spin correlation \cite{Stelzer06}:
\begin{equation}
C = \frac{\sigma(t_L\bar{t}_L + t_R\bar{t}_R) -
\sigma(t_L\bar{t}_R + t_R\bar{t}_L)}{\sigma(t_L\bar{t}_L + t_R\bar{t}_R) +
\sigma(t_L\bar{t}_R + t_R\bar{t}_L)} \,.
\label{eq:c}
\end{equation}
This quantity depends on the spin quantization axis.
At leading order,
the SM prediction is $C= -0.471$
for the helicity basis which we choose in this work.
It is known that NLO correction to $C$ is rather large, shifting
$C$ to $-0.352$~\cite{Stelzer06}.
In Ref.~\cite{Jung:2009pi}, 
we propose a new spin-spin FB asymmetry $C_{FB}$  defined as
\begin{equation}
C_{FB} \equiv  C (\cos\theta \geq 0) -  C (\cos\theta \leq 0)  ,
\end{equation}
where $C(\cos\theta \geq 0 (\leq 0))$ implies that the cross sections in the
numerator of Eq.~(7) are obtained for the forward (backward) region:
$\cos\theta \geq 0 (\leq 0)$.
This quantity can be measured by dividing the $t\bar{t}$ sample into
the forward top and the backward top events.

In Ref.~\cite{Jung:2010yn}, we note that the NP interaction
responsible for the deviation of $A_{\rm FB}$ from the SM prediction 
is parity($P$)-violating.  Motivated by the observation, we propose
new $P$-odd observables:
\begin{eqnarray}
D &\equiv & \frac{\sigma(t_R\bar{t}_L) - \sigma(t_L\bar{t}_R)}
{\sigma(t_R\bar{t}_R) + \sigma(t_L\bar{t}_L) +
\sigma(t_L\bar{t}_R) + \sigma(t_R\bar{t}_L)}\,,
\nonumber \\[0.1cm]
D_{\rm FB} &\equiv &
D (\cos\hat\theta \geq 0) -  D (\cos\hat\theta \leq 0)
\end{eqnarray}
which correspond to the difference between the 
longitudinal polarizations of top and antitop quarks. 
We show the $P$-odd new observables provide important information 
on the chiral structures of
NP that might be relevant to the $A_{\rm FB}$.

The leading-order SM predictions for the 
proposed new observables are $C_{\rm FB}=D=D_{\rm FB}=0$.
We note that, being different from
$C_{\rm FB}$, the observables $D$ and $D_{\rm FB}$ vanish in QCD
to all orders because of $P$ conservation. But, in the presence of NP,
we still need to know how much changes will be induced
by the NP$+$QCD corrections for quantitative studies.

\section{Flavor-conserving Case}
\label{sec:fc}

Firstly, we consider the FC couplings $g^{t}_{L,R}$ only 
for the connection to $A_{\rm FB}$ taking $m_V=150$ GeV. 
Without losing generality, $g^{t}_{L}=g^{u}_{L}$ is taken.

With $ \sigma(p\bar{p}\to W^\pm V)
=K\,\sigma_{LO}(p\bar{p}\to W^\pm V) $ taking $K=1.3$, one may obtain
\begin{equation}
{\rm {\bf CDF}~ellipse}~:~(g^u_L)^2+(g^d_L)^2-1.2\, g^u_L g^d_L 
= 0.060\,\left(\frac{\sigma(p\bar{p}\to W^\pm V)}{4\,{\rm pb}}\right)\,.
\end{equation}
If the dijet excess reported by CDF diminishes, then the CDF ellipse should 
shrink accordingly, as described by the above equation.
On the other hand, with $ \sigma(p\bar{p}\to V)
=K\,\sigma_{LO}(p\bar{p}\to V) $ taking $K=1.3$, we get
\begin{equation}
{\rm {\bf UA2}~ellipse}~:~\left[(g^u_L)^2+0.18(g^d_L)^2\right]+
\left[(g^u_R)^2+0.18(g^d_R)^2\right] \lsim 0.075\,
\left(\frac{\sigma^{\rm max}(p\bar{p}\to V)}{300\,{\rm pb}}\right)\,.
\end{equation}

\begin{figure}[!t]
\begin{center}
{\epsfig{figure=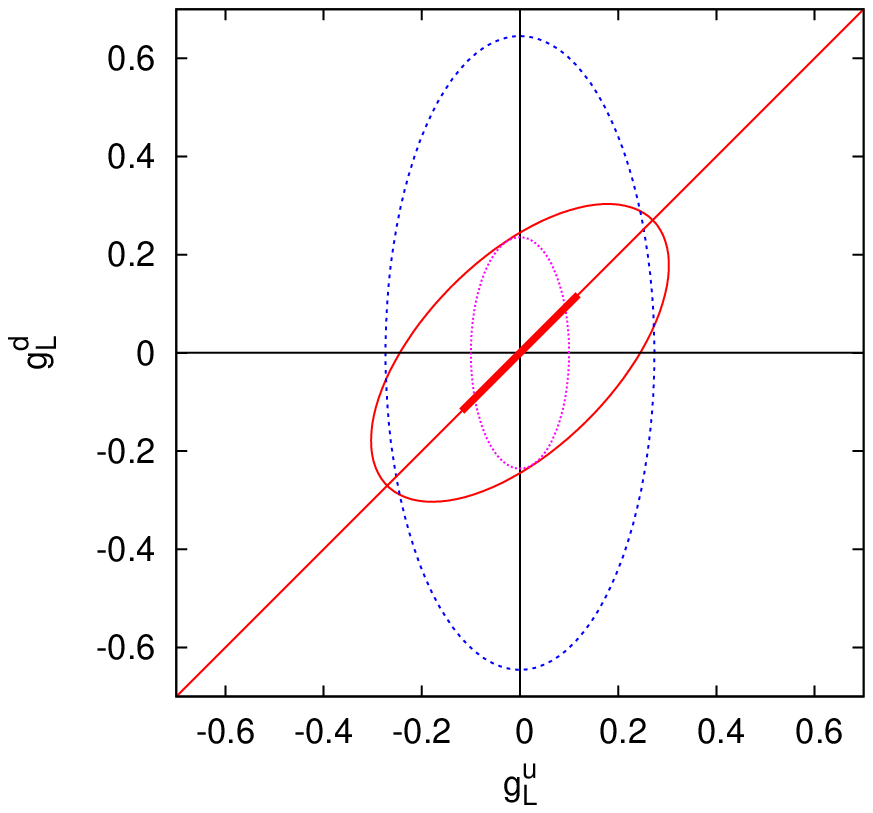,height=7.0cm,width=8.5cm}}
\hspace{-1.5cm}
{\epsfig{figure=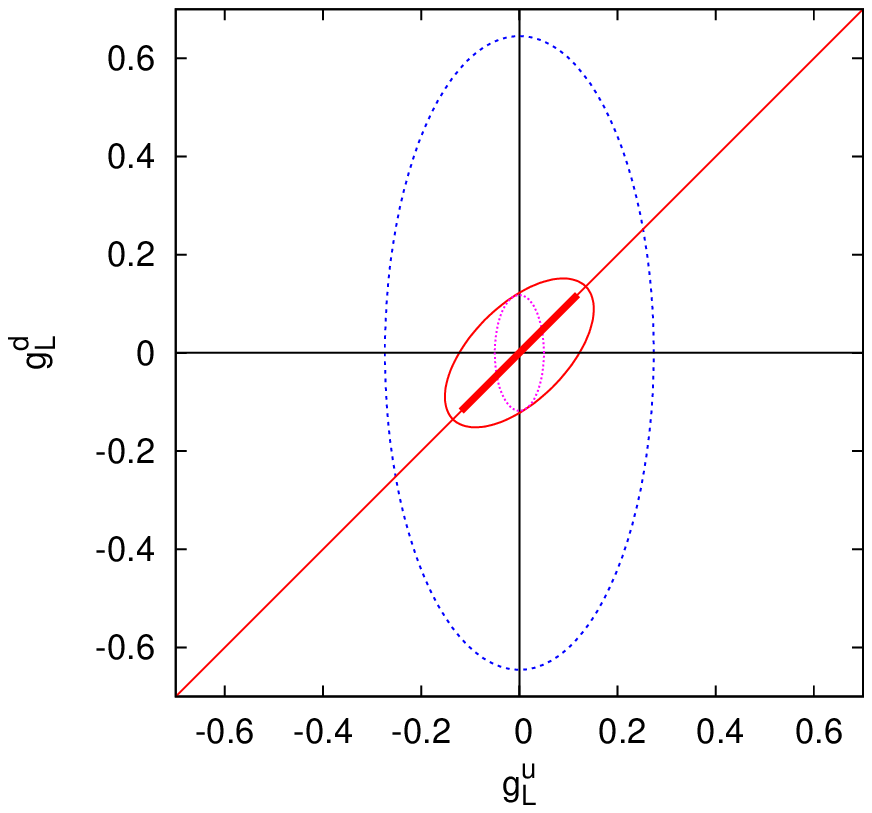,height=7.0cm,width=8.5cm}}
\end{center}
\vspace{-1.0cm}
\caption{\it The {\bf CDF} (tilted red) and {\bf UA2} (co-axial inner and outer)
ellipses on the $g^u_L$-$g^d_L$ plane taking
$m_V=150$ GeV and  $\sigma^{\rm max}(p\bar{p}\to V)=300$ pb
with 
$\sigma(p\bar{p}\to W^\pm V)=4$ pb
(left) and $1$ pb (right).
See text for explanation.
}
\label{fig:guL-gdL}
\end{figure}

In Fig.~\ref{fig:guL-gdL}, we show the {\bf CDF} (tilted red) and 
the {\bf UA2} (co-axial inner and outer) ellipses on the $g^u_L$-$g^d_L$ 
plane, taking $\sigma(p\bar{p}\to W^\pm V)=4$ pb (left)
and $1$ pb (right) with
$\sigma^{\rm max}(p\bar{p}\to V)=300$ pb.
For both cases, the outer {\bf UA2} (blue) ellipses
are obtained by taking $g^u_R=g^d_R=0$ while the inner (magenta) ones are 
for $(g^u_R)^2+0.18(g^d_R)^2 = \left[(g^u_R)^2
+0.18(g^d_R)^2\right]_{\rm max} \simeq 0.065$ (left) and $0.072$ (right).
If the sum of the right-handed couplings squared is larger than $0.065$, there is
no solution for the  CDF dijet excess corresponding to $\sigma (WV) = 4$ pb.
When the sum increases further, the {\bf UA2} bound becomes stronger.
And, if it is larger than $0.072$, the
left-handed couplings are too small to achieve even $\sigma (WV) = 1$ pb.
Therefore, $(g^u_R)^2+0.18(g^d_R)^2 \lsim 0.072$ is required to accommodate
$\sigma (WV) = 1 - 4 $ pb.

The thick solid (red) straight line along $g^u_L=g^d_L$ in Fig.~\ref{fig:guL-gdL}
presents the case of the leptophobic $Z^\prime$ model in a particular 
type of $E_6$ model 
~\cite{Rosner:1996eb}, where  the couplings of $Z^\prime$ to the SM quarks are 
related as $g^u_L=g^d_L=g^u_R/2=-g^d_R$.
We find the leptophobic model gives at most
$\sigma(p\bar{p}\to W^\pm V) \simeq 0.77 $ pb, which is rather small compared with 
the presumed cross section of $1 - 4$ pb.

\begin{figure}[!t]
\begin{center}
{\epsfig{figure=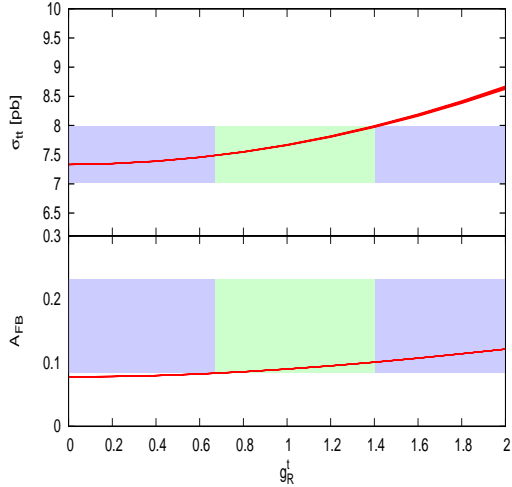,height=7.0cm,width=7.0cm}}
\end{center}
\vspace{-1.0cm}
\caption{\it The $t\bar{t}$ production cross section and 
the forward-backward asymmetry are shown as functions of
the coupling $g^t_R$ in the FC case 
taking $m_V=150$ GeV and $(g^u_R)^2+0.18(g^d_R)^2 =
\left[(g^u_R)^2+0.18(g^d_R)^2\right]_{\rm max} \simeq 0.065$ with
$g^u_R$ varying from $0$ to $(g^u_R)_{\rm max}\sim 0.25$. The left-handed couplings
are given by $(g^u_L,g^d_L)\simeq (\pm 0.029\,,\mp 0.23)$ and
we have taken $g^t_L=g^u_L$.
The two horizontal bands show the experimental $1$-$\sigma$ regions,
$\sigma_{t\bar{t}}=7.50\pm 0.48$ pb and
$A_{\rm FB}=0.158\pm 0.074$~\cite{Aaltonen:2011kc}.
The vertical bands show the $1$-$\sigma$
allowed region for the coupling $g^t_R$.
}
\label{fig:case2-3}
\end{figure}

When $g^u_R=g^d_R=0$, we find the solutions, represented by the points
on the tilted {\bf CDF} ellipse inside of the outer {\bf UA2} one,
always result in $A_{\rm FB} < A_{\rm FB}^{\rm SM}$, which is not satisfactory.
Therefore, one cannot explain both the top FB asymmetry and the CDF dijet excess 
in terms of FC: $g_L^u = g_L^t \neq 0$ and $g_R^u = g_R^d = 0$.

On the other hand, when 
the sum $(g^u_R)^2+0.18(g^d_R)^2$ takes its maximum value $\simeq 0.065$
with $\sigma(p\bar{p}\to W^\pm V) = 4$ pb,
we have the solutions $(g^u_L,g^d_L)\simeq (\pm 0.029\,,\mp 0.23)$
given by the two overlapping points of the {\bf CDF} and 
inner {\bf UA2} ellipses, see the left frame of Fig.~\ref{fig:guL-gdL}. 
In this case,  we find the simultaneous solutions to the CDF dijet 
excess and the large $A_{\rm FB}$ 
are possible if $0.67 \lsim g^t_R \lsim 1.4$,  see Fig.~\ref{fig:case2-3}.
We observe this is quite large compared to $(g^u_R)_{\rm max} \sim 0.25$.
Note that the coupling $g^t_R \sim 1.4$ is still small enough for 
perturbation, since $(g_s g^t_R )^2/4 \pi \sim 0.2$.

The $t\bar{t}$ production cross section and
the forward-backward asymmetry are not much affected 
by taking the smaller $\sigma(p\bar{p}\to W^\pm V) = 1$ pb.
This is because the NP contribution to the top-quark pair production 
is dominated by the right-handed couplings and their squared sum
increases only by $\sim 10$ \%.

\section{Flavor-violating Case}
\label{sec:fv}

Secondly, taking $m_V=150$ GeV again, 
we consider the right-handed FV coupling $\tilde{g}^t_R$
with $\tilde{g}^t_L=g^t_R=0$
for the connection to $A_{\rm FB}$
while keeping the left-handed FC couplings $g^t_L$.

In this case the {\bf UA2} ellipses remain the same
but the {\bf CDF} one has an additional term:
\begin{equation}
{\rm {\bf CDF}~ellipse}~:~(g^u_L)^2+(g^d_L)^2-1.2\, g^u_L g^d_L 
+0.0070\,|\tilde{g}^t_R|^2 = 
0.060\,\left(\frac{\sigma(p\bar{p}\to W^\pm V)}{4\,{\rm pb}}\right)\,.
\end{equation}
To reduce the number of independent couplings, we assume the relations
\begin{equation}
g^u_L=g^t_L=g^d_L/\omega_L\,, \ \ \ g^u_R=g^d_R\equiv \omega_R\,g^u_L\,,
\label{eq:case3-1}
\end{equation}
with $\omega_L^2= 1$. Under the assumption, 
by combining the {\bf UA2} and {\bf CDF} conditions,
one may obtain an inequality
\begin{equation}
\sigma(p\bar{p}\to W^\pm V)/{\rm pb} \lsim (2-1.2\, \omega_L)
\left[\frac{4.2}{1+\omega_R^2}\,
\left(\frac{\sigma^{\rm max}(p\bar{p}\to V)}{300\,{\rm pb}}\right)\right]
\ + \
0.47\,|\tilde{g}^t_R|^2 \,.
\label{eq:swvmax}
\end{equation}
We observe the cross section becomes smaller with the choice $\omega_L=+1$
as $\omega_R$ grows.

\begin{figure}[!t]
\begin{center}
{\epsfig{figure=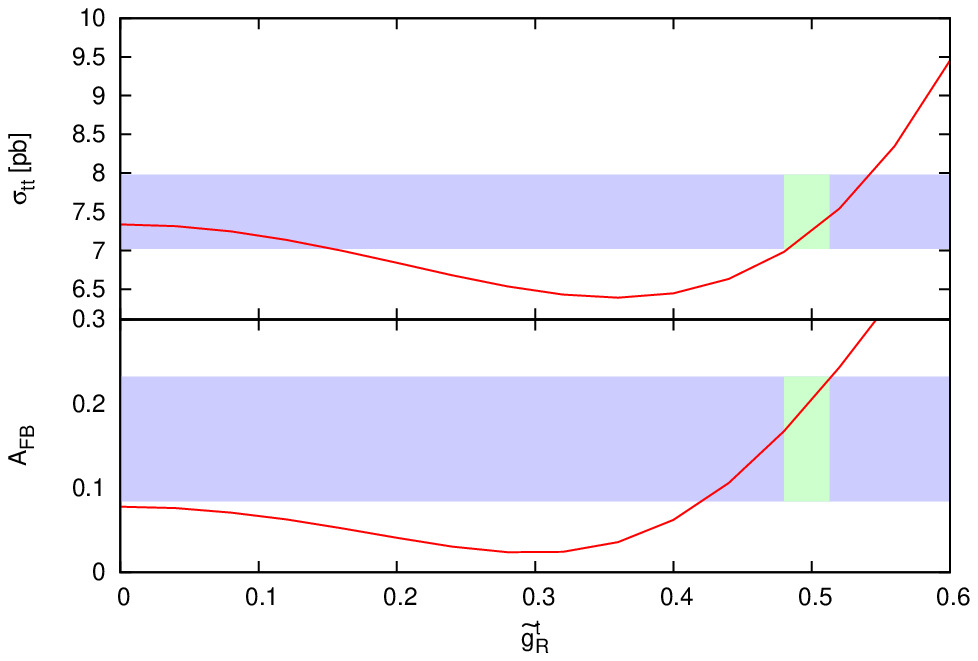,height=7.0cm,width=7.0cm}}
{\epsfig{figure=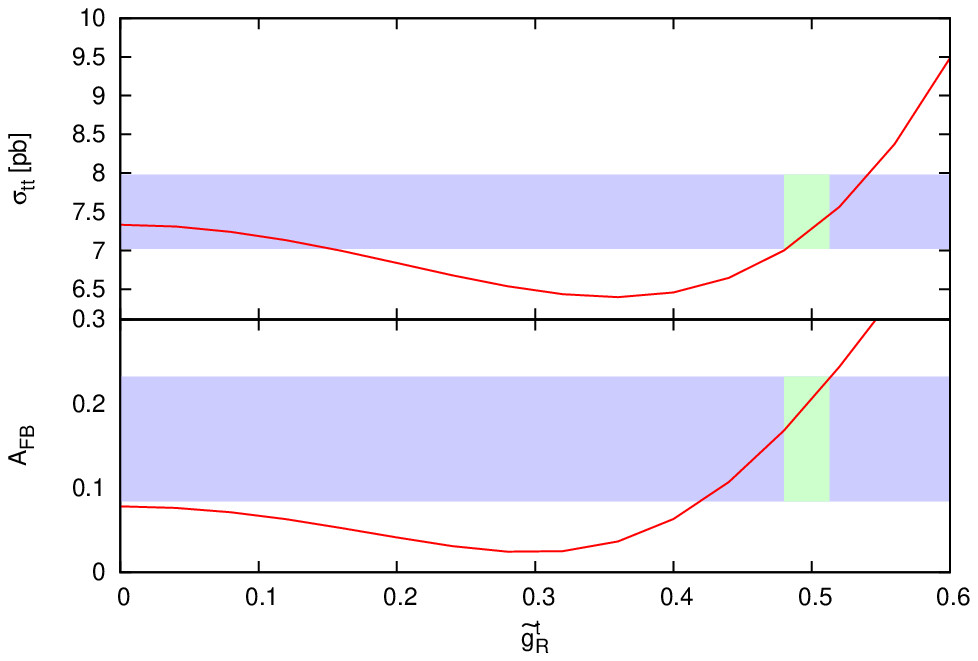,height=7.0cm,width=7.0cm}}
\end{center}
\vspace{-1.0cm}
\caption{\it The $t\bar{t}$ production
cross section and the forward-backward asymmetry  in the FV case
as functions the coupling $\tilde{g}^t_R$ taking $m_V=150$ GeV.
Left: The other couplings are given by Eqs.~(\ref{eq:case3-1}) and
(\ref{eq:case3-2}) with $\omega_L=-1$ and $g^t_R=\tilde{g}^t_L=0$.
Right: The only non-vanishing coupling other than $\tilde{g}^t_R$
is $g^u_L$ which takes a more or less fixed value of 
about $~0.24\sqrt{\sigma(p\bar{p}\to W^\pm V)/4\,{\rm pb}}$.
The vertical bands show the $1$-$\sigma$
allowed region for the coupling $\tilde{g}^t_R$.
}
\label{fig:case3}
\end{figure}

Taking $\omega_L=-1$ and
$\sigma^{\rm max}(p\bar{p}\to V)=300$ pb, we have
\begin{eqnarray}
g^u_L &=& \sqrt{\frac{\left[0.060
\left(\frac{\sigma(p\bar{p}\to W^\pm V)}{4\,{\rm pb}}\right)
-0.0070\,|\tilde{g}^t_R|^2\right]}{3.2}}\,; \ \ 
g^u_R = \sqrt{\frac{0.075-1.2\, (g^u_L)^2}{1.2}}
\label{eq:case3-2}
\end{eqnarray}
where $g^u_R$ is fixed to saturate the {\bf UA2} bound. 
When $\sigma(p\bar{p}\to W^\pm V)=4$ pb,
we find the simultaneous solutions to the CDF dijet 
excess and the large $A_{\rm FB}$ 
are possible if 
$0.48 \lsim \tilde{g}^t_R \lsim 0.51$,
see the left frame of Fig.~\ref{fig:case3}, giving
$g^u_L \sim 0.13$ and $g^u_R \sim 0.21$.
%
When $\sigma(p\bar{p}\to W^\pm V)=1$ pb, 
while we have the smaller coupling $g^u_L \sim 0.064$ 
with $g^u_R \sim 0.24$, we find
that the range of $\tilde{g}^t_R$ for the simultaneous 
solutions almost remains the same since
the NP contribution to the top-quark production is dominated
by the $t$-channel diagram which depends only on
the coupling $\tilde{g}^t_R$.

Taking $\omega_L=+1$ and $\sigma^{\rm max}(p\bar{p}\to V)=300$ pb, we find
$\sigma(p\bar{p}\to W^\pm V)$ can not be larger than
$\sim (3.4/(1+\omega_R^2) + 0.47\,|\tilde{g}^t_R|^2)$ pb, see Eq.~(\ref{eq:swvmax}),
in which we find $|\tilde{g}^t_R| \lsim 0.5$ constrained by
the $t\bar{t}$ production cross section. 
Therefore, the right-handed couplings are constrained by
$\omega_R^2 \lsim 2.8$ in order to have $\sigma(p\bar{p}\to W^\pm V) \gsim 1$ pb.

Actually, in the FV case, we find that it is possible  to explain the
CDF dijet excess and the large $A_{\rm FB}$ simultaneously only with
the two couplings $g^u_L$ and $\tilde{g}^t_R$
by noting that the $s$-channel contribution to the
$t\bar{t}$ production is negligible under the assumption $g^t_L=g^u_L$
(\ref{eq:case3-1}).
Only with $g^u_L$ and $\tilde{g}^t_R$ non-vanishing, we observe that
$g^u_L$ takes a more or less fixed value of about
$0.24\sqrt{\sigma(p\bar{p}\to W^\pm V)/4\,{\rm pb}}$
and $\tilde{g}^t_R$ should take values between $0.48$ and $0.51$,
see the right frame of Fig.~\ref{fig:case3} 
which does not show any visible difference from the left frame
reflecting the negligible $s$-channel contribution via 
the FC couping $g^t_L$.

\begin{figure}[!t]
\begin{center}
\hspace{-1.0cm}
{\epsfig{figure=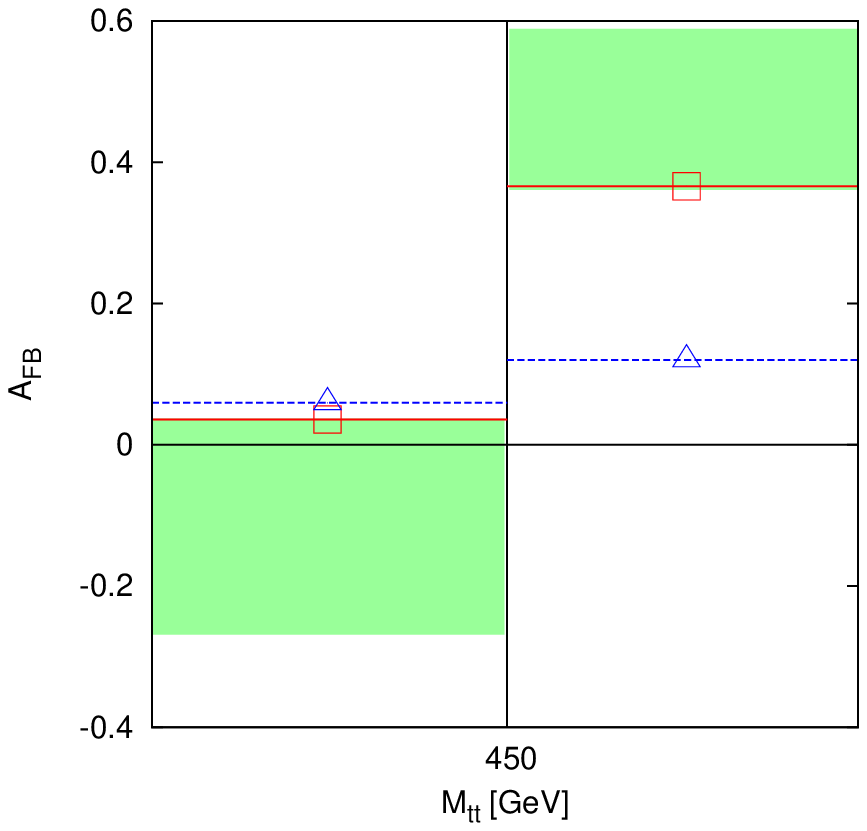,height=5.5cm,width=6.5cm}}
\hspace{-1.5cm}
{\epsfig{figure=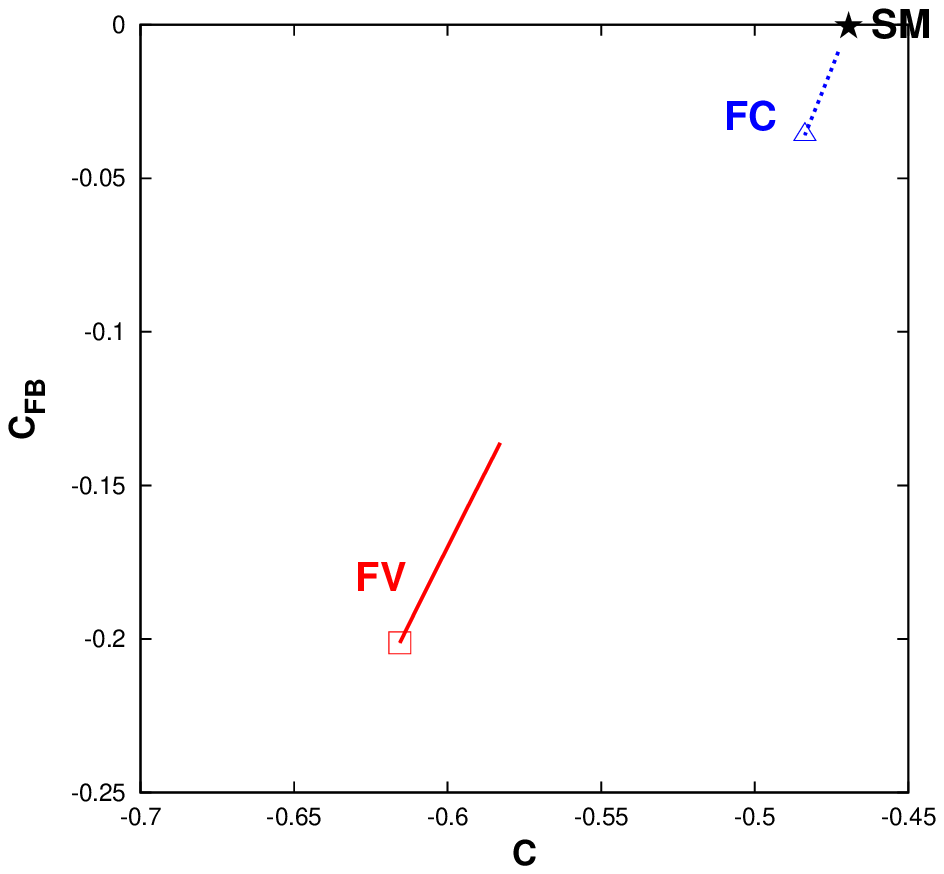,height=5.5cm,width=6.5cm}}
\hspace{-1.5cm}
{\epsfig{figure=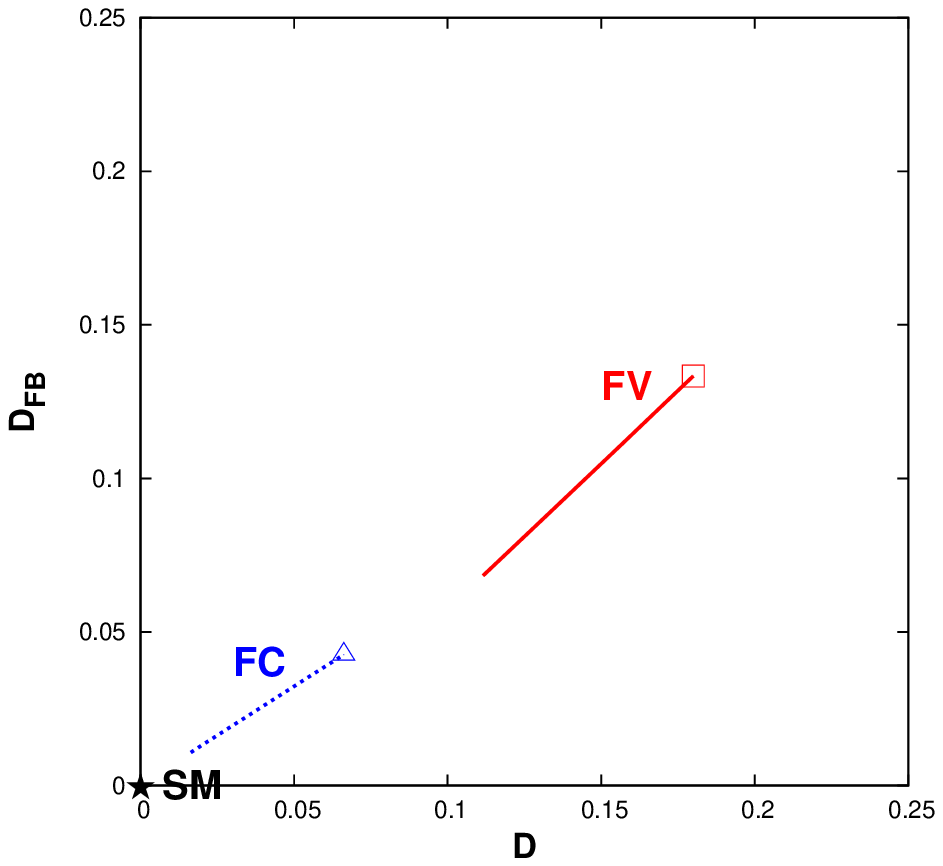,height=5.5cm,width=6.5cm}}
\end{center}
\vspace{-0.5cm}
\caption{\it 
Left: The $M_{t\bar{t}}$ dependent $A_{\rm FB}$ for
the FC point with $g^t_R=1.4$ (triangle)
and the FV points with $\tilde{g}^t_R=0.51$ (square).
Middle and Right: The polarization observables
$C$ and $C_{\rm FB}$ (middle) and
$D$ and $D_{\rm FB}$ (right) in the
$1$-$\sigma$ allowed regions for the 
FC and FV couplings: 
$0.67 \protect\lsim g^t_R \protect\lsim 1.4$ and 
$0.48 \protect\lsim \tilde{g}^t_R \protect\lsim 0.51$.
The leading order SM predictions are denoted by stars.
}
\label{fig:mtt}
\end{figure}

Finally, in the left frame of Fig.~\ref{fig:mtt}, 
we show $A_{\rm FB}$ in the low and high 
invariant mass regions of the top-quark pair 
taking  the FC point with $g^t_R=1.4$ (triangle)
from Fig.~\ref{fig:case2-3}
and the FV points with $\tilde{g}^t_R=0.51$ (square) from
Fig.~\ref{fig:case3}.
We observe that the FV case leads to the
consistent results with the current measurement of the
mass dependent FB asymmetry~\cite{Aaltonen:2011kc}.
For the FC case, we have somewhat lower $A_{\rm FB}$ than the current data 
and the future analysis with more
data could tell more definitely whether the FC case is viable or not.

In the middle and right frames of Fig.~\ref{fig:mtt},
we show our predictions for the polarization observables 
$C$ and $C_{\rm FB}$ and
$D$ and $D_{\rm FB}$, respectively, in the
$1$-$\sigma$ allowed regions for the 
FC coupling $g^t_R$ and
FV coupling $\tilde{g}^t_R$.
We find $-C$ and $-C_{\rm FB}$ can be as large as $0.62$ and $0.2$, 
respectively, in the FV case. On the other hand,
$D$ and $D_{\rm FB}$ can be as large as $0.18$ and $0.13$, respectively.
The SM prediction is $C=-0.352$ at NLO and, for other observables,
the leading-order SM predictions are $C_{\rm FB}=D=D_{\rm FB}=0$.
To our best knowledge, there are no available
calculations of  the latter
including QCD corrections which renders
it difficult to make a decisive model discrimination at the current stage.
Implementing QCD corrections to these
observables deserves more theoretical works in the future.
With more data accumulated at the Tevatron and LHC, these
new observables of $C_{\rm FB}$, $D$, and $D_{\rm FB}$ can 
give useful and independent information on NP scenarios.

Before we close this section we comment on the {\em direct} constraint
on the flavor-changing coupling on $\tilde{g}^t_R$ obtained
by the CMS Collaboration in search for same-sign top-quark pair production
at the LHC~\cite{Chatrchyan:2011dk}. 
The constraint is $\tilde{g}^t_R \lsim 0.4$ when $m_V=150$ GeV
at the 95 \%  confidence level.
Taking this limit seriously, our FV solution  for $A_{\rm FB}$
with $\tilde{g}^t_R \sim 0.5$ is not viable.

\section{Conclusions}
\label{sec:conclusions}

In this work, we have introduced a neutral color-singlet vector boson $V_{\mu}$
in order to explain the CDF dijet excess. We find that each of the exchanges of
the left-handed $u$ and $d$ quarks can easily accommodate 
$\sigma(p\bar{p}\to W^\pm V) \sim 1-4 $ pb,  if $m_V=150$ GeV
with $g^u_L\,,g^d_L \sim 0.12 - 0.24$. When $g^u_L=-g^d_L\sim 0.064 - 0.13$, we may
also have the $1-4$  pb cross section. 
On the other hand, with
$g^u_L=+g^d_L$, the cross section $\sigma(p\bar{p}\to W^\pm V)$
becomes smaller due to the destructive 
interference between the $t$- and $u$-channel diagrams 
combined with the UA2 bounds and the right-handed couplings are constrained
by $(g^{u,d}_R)^2 \lsim 2.8\, (g^{u,d}_L)^2 $ 
in order to have $\sigma(p\bar{p}\to W^\pm V) \gsim 1$ pb.

Towards the simultaneous explanation for both the top FB asymmetry 
and the dijet resonance, we consider the FC and FV couplings of the top 
quarks to the same vector boson relevant to the CDF dijet excess.
In the FC case, we find the two puzzles can be resolved with
sizeable right-handed couplings of $g^u_R \sim 0.25$ or $g^d_R \sim 0.6$ 
together with $0.67 \lsim g^t_R \lsim 1.4$. 
This solution is almost independent of $\sigma(p\bar{p}\to W^\pm V)$ because
the NP contribution to the top-quark pair production 
is dominated by the right-handed couplings.
In the FV case, the coupling
$\tilde{g}^t_R \sim 0.5$ may provide the simultaneous solutions
with $g^u_L=-g^d_L\sim 0.064-0.13$ or $g^u_L \sim 0.12-0.24$
to accommodate $ 1~{\rm pb} \lsim \sigma(p\bar{p}\to W^\pm V) \lsim 4 $ pb.
Again, we observe that the solution for $A_{\rm FB}$ with $\tilde{g}^t_R \sim 0.5$
is almost independent of $\sigma(p\bar{p}\to W^\pm V)$ because
the NP contribution to the top-quark production is dominated
by the $t$-channel diagram which depends only on
the coupling $\tilde{g}^t_R$.

From Fig.~5, we observe that the FC and FV cases 
can be distinguished in the mass dependence of 
the top FB asymmetry if more data is accumulated and analyzed. 
However one can have additional handles to diagnose the new physics 
structure using the FB spin-spin correlation or longitudinal top polarization,
as suggested in Refs.~\cite{Jung:2009pi,Jung:2010yn}.  
It is highly desirable to measure and extract these observables from the current 
data set obtained at the Tevatron. 

Having identified the coupling structures that are needed to accommodate 
both the CDF dijet excess and the top FB asymmetry, the next important question 
would to construct a realistic model with such a neutral vector boson. 
Since the couplings are flavor dependent and might have a large FV
$u_R - t_R$ coupling, it might be challenging to build such a model. 
(See, for example, Refs.~\cite{Nelson:2011us,Jung:2011zv,Jung:2011ua,Ko:2011vd} 
for the attempts in this direction.)

Independently of the model building, a light vector boson ($\lsim 200$ GeV) 
coupling only to quarks has been elusive at colliders. The $Wjj$ channel at 
hadron colliders is one of the best to probe such light leptophobic gauge boson.
It remains to be seen if the further data analysis at the Tevatron and the LHC
could give any hint for or strong constraint on such a leptophobic gauge boson. 

There are some more phenomenological issues: $(i)$ the LHC signatures of the
FC and FV solutions found in this work
\footnote{
At the LHC with 7 TeV, taking $m_V=150$ GeV, we find that 
$\sigma_{LO}(pp\to W^\pm V) \sim 22$ pb and $24$ pb 
for the FC and FV cases, respectively,
see the right frame of Fig.~\ref{fig:qqwv}.},
$(ii)$ the production of the NP particle ($V$)
associated with $\gamma$, $W^\pm$, $Z$, and $V$ itself at hadron colliders,
$(iii)$ the single top and same-sign top pair productions,
$(iv)$ other possibilities with neutral and charged NP particles of  
color-singlet/octet scalar and vector bosons, 
$(v)$ model discrimination by use of the top-quark polarizations, 
etc.  We address these issues in future publications.

\subsection*{Acknowledgements}
The work by PK is supported in part by Korea National Research Foundation 
through Korea Neutrino Research Center (KNRC) at Seoul National University.


\end{document}